# [1]p/π+ response of single layer THGEM detector in Ar+3% iC$_4$H$_{10}$

Hong Daojin (洪道金)[1,2,3;1] YU Boxiang (俞伯祥)[3,4] Liu Hongbang （刘宏邦）[1,2;2] He Xiaorong (何小荣)[1]
An Guangpeng (安广朋)[3,4] Chen haitao (陈海涛)[3,4] Chen Shi (陈石)[2] Hu tao (胡涛)[3,4]
Li Jia-cai （李家才）[3,4] Liu Qian (刘倩)[2] Niu Shunli (牛顺利)[3,4] Ruan Xiangdong (阮向东)[1]
Xie yigang (谢一冈)[2,4] Zhang xuan (张烜)[3,4] Zheng yangheng (郑阳恒)[2]

1 Department of Physics, Guangxi University, Nanning 530004, China
2 University of Chinese Academy of Sciences, Beijing 100049, China
3 State Key Laboratory of Particle Detection and Electronics, Beijing 100049, China
4 Institute of High Energy Physics, Chinese Academy of Sciences, Beijing 100049, China

**Abstract**: In this work, we study the response of a single layer Thick Gaseous Electron Multiplier (THGEM) detector to p/π+ at the E3 line of the Beijing Test Beam Facility. The THGEM detector drift gap used in this study is 4 mm, and the gain of the detector operated in Ar+3% iC$_4$H$_{10}$ is about 2000. p/π+ particles at momenta between 500 MeV/c and 1000 MeV/c are distinguished by a Time Of Flight (TOF) system. Results show that at the measured momenta, detection efficiencies for p are from 93% to 99%, and for π+ in the range of 82%~88%. Meanwhile, simple Geant4 simulations also have been done, and are consistent with the amplitude spectra of beam test results. We preliminarily study the feasibility of the THGEM detector as sampling element for a Digital Hadronic Calorimeter (DHCAL), which may provide support for a THGEM detector possibly being applied in the Circular Electron Positron Collider (CEPC) HCAL.

## 1 Introduction

In recent years, THGEM (Thick Gaseous Electron Multiplier) detectors have been studied extensively. It is a simple, cheap and robust detector, offering a rise time of a few ns and sub mm spatial resolution. Due to its good performance, the THGEM detector has many meaningful applications. It is not only used in X-ray diffraction [1] or X-ray imaging [2], but also as a cosmic-ray muon hodoscope [3] and as a UV-photon detector for RICH [4].

1∗ Supported by National Natural Science Foundation of China (11265003, U1431109, 11205240), State Key Laboratory of Particle Detection and Electronics.

1) E-mail: hongdj@ihep.ac.cn
2) E-mail: liuhb@gxu.edu.cn

Currently, the Circular Electron Positron Collider (CEPC) Higgs factory is being proposed for the purpose of precision measurements of the Higgs particle, and the Digital Hadronic Calorimeter (DHCAL) will play a crucial role in this research. As one of the DHCAL candidates [5], THGEM/GEM is attractive due to its above-mentioned properties. A large area THGEM detector could be considered for the DHCAL sampling element. Moreover, there have already been studies on THGEM-based DHCAL internationally [6, 7], but related research has rarely been done in China [8].

This paper reports preliminary beam test results at the E3 line of the Beijing Test Beam Facility, aimed at studying the response of a single layer THGEM detector to $p/\pi^+$ at different momenta. A THGEM detector with 4 mm drift gap and 2 mm induction gap operated in $Ar+3\%\ iC_4H_{10}$ has been extensively studied in the lab [9], so a THGEM detector under these conditions was selected as the research object. Based on the whole results including amplitude spectra, $p/\pi^+$ detection efficiencies and Geant4 simulation results, applicability of the THGEM detector as a DHCAL sampling element may be expected.

## 2 Experimental setup

### 2.1 Experimental setup

The beam test was conducted at the E3 line of the Beijing Test Beam Facility, which is generated by shooting a target with 2.5 GeV/c electron beam. It provides a mixed beam including e, $\pi^+$ and p, and the maximum particle counting rate is about 3 Hz. The electrons are identified by a Cherenkov detector, and $p/\pi^+$ are distinguished by their different in flight time with a TOF system. In our experiment, mainly p and $\pi^+$ are measured.

The simple setup of the beam test is shown in Fig. 1. There are three plastic scintillator detectors (TOF1, TOF2 and TOF3) on the beam line, and the THGEM detector is located between TOF2 (5x5 cm$^2$) and TOF3 (3x3 cm$^2$). In this work, TOF1 (5x5 cm$^2$) and TOF3 are used to provide a trigger for a CAEN DT5751. With TOF2 added for a triple coincidence, the real $p/\pi^+$ events can be further purified by reducing fake events.

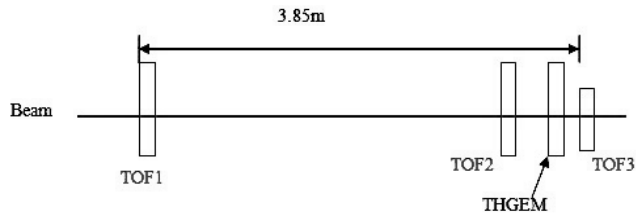

Fig.1. Experimental setup

Fig. 2 shows the electronics system of this experiment. The THGEM detector signal is read out on a single pad combined with a charge sensitive preamplifier (ORTEC 142AH) followed by a linear shaping amplifier (CAEN N968) with 0.5 μs shaping time, coarse gain of 50 and fine gain of 1. Once the DT5751 is triggered by the coincidence signal, the signals from the THGEM detector and TOF detectors are recorded by the DT5751 data acquisition program of, thus acquiring the waveforms of the signals from the four channels.

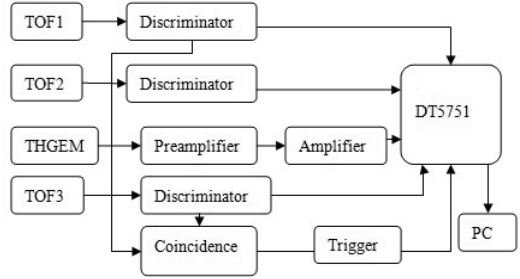

Fig.2. Electronics system

The flight time of p/$\pi^+$ is set as the difference between TOF1 and TOF3, the distance between TOF1 and TOF3 being about 3.85 m for this test. According to relativity theory, the flight time of p and $\pi+$ at different momenta can be calculated as shown in Table 1, from which p and $\pi^+$ can be distinguished clearly.

Table 1.Flight time of p/$\pi^+$ at a distance of 3.85 m

| Momentum | $\pi^+$(139.657MeV/c$^2$) | | p(938.279MeV/c$^2$) | |
| --- | --- | --- | --- | --- |
| MeV/c | E(MeV) | t(ns) | E(MeV) | t(ns) |
| 500 | 519.1 | 13.289 | 1063.2 | 27.217 |
| 600 | 616.0 | 13.142 | 1113.7 | 23.761 |
| 700 | 713.8 | 13.052 | 1170.6 | 21.405 |
| 800 | 812.1 | 12.994 | 1233.0 | 19.729 |
| 900 | 910.8 | 12.954 | 1300.1 | 18.489 |
| 1000 | 1009.7 | 12.924 | 1371.3 | 17.554 |

## 2.2 The THGEM detector

The single layer THGEM used in this study is 5x5cm$^2$ in area and 0.2 mm thick, with the holes having 0.5 mm pitch, 0.2 mm diameter and 10 μm rim. The detector, which has 4 mm drift gap and 2 mm induction gap, is operated in Ar+3% iC$_4$H$_{10}$. A schematic of the THGEM detector is shown in Fig. 3. The electric fields $E_{drift}$ and $E_{ind}$ of the THGEM detector combined with a preamplifier followed by a linear shaping amplifier are set to be 640 V/cm and 2 kV/cm. The THGEM voltage is 545 V, thus the gain of the THGEM detector is approximately 2000 [10], which is sufficient for this test, and which also avoids high voltage discharge in the detector.

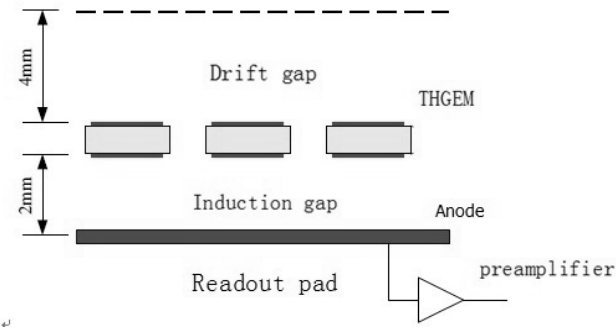

Fig.3. Schematic of the THGEM detector

**2.3 Measurement of baseline and detector stability**

To test the response of the detector without beam condition, the baseline of the THGEM detector has been measured with a random trigger. Fig. 4 shows the results; the data can be fitted with a Gaussian function, with the mean and sigma of the Gaussian distribution being about 180 mV and 5 mV respectively. A threshold value for the signal could be set at the mean plus 3 times sigma, so if the THGEM detector signal is higher than 195 mV, it is considered a valid event.

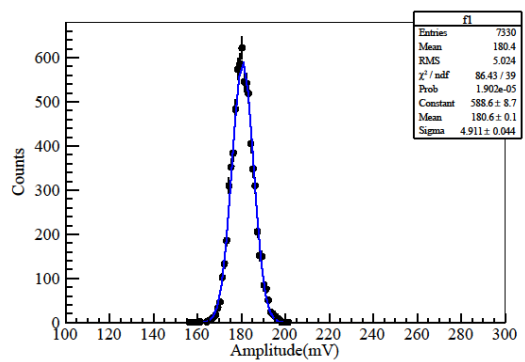

Fig.4. Baseline measurement with a random trigger

We also check the gain of the THGEM detector by calibration with $^{55}$Fe. The $^{55}$Fe signal is attenuated by 3 dB before being fed into the DT5751, with the result shown in Fig. 5. Factors influencing the THGEM detector's performance (environmental conditions, starting gas flow and electronics factors etc) are unstable before 30hours. As time goes on, they gradually stabilize under beam conditions.

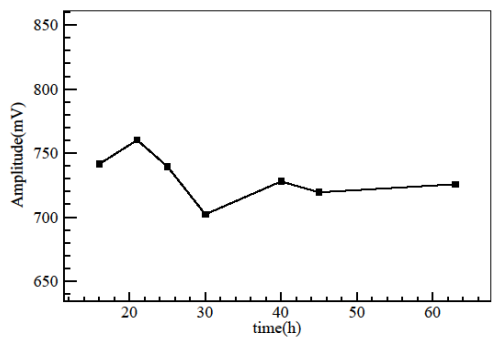

Fig.5. Measurement of detector stability calibrated by $^{55}$Fe

# 3 Results

## 3.1 Geant4 simulation results

A 4 mm drift gap filled with Ar+3% $iC_4H_{10}$ has been simulated with Geant4, and Fig. 6 shows the simulation results. The deposition energy of p is between 0.7 keV and 1.8 keV, and is 1-4 times higher than that for $\pi^+$. Besides, the deposition energy of p decreases as the momentum increases, but for $\pi^+$ it is almost constant.

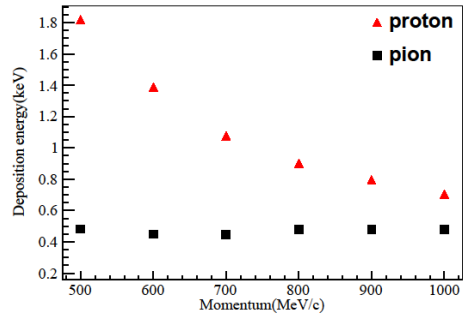

Fig.6. Geant4 simulation of $p/\pi^+$ deposition energy in 4 mm Ar+3% $iC_4H_{10}$

## 3.2 Amplitude spectrum of $p/\pi^+$

Fig. 7 shows the flight time of $p/\pi^+$ at different momenta, with the flight time of p larger than that of $\pi^+$. Comparing with the expected values from Table 1, the flight times of $p/\pi^+$ are reasonable, thus $p/\pi^+$ can be distinguished clearly.

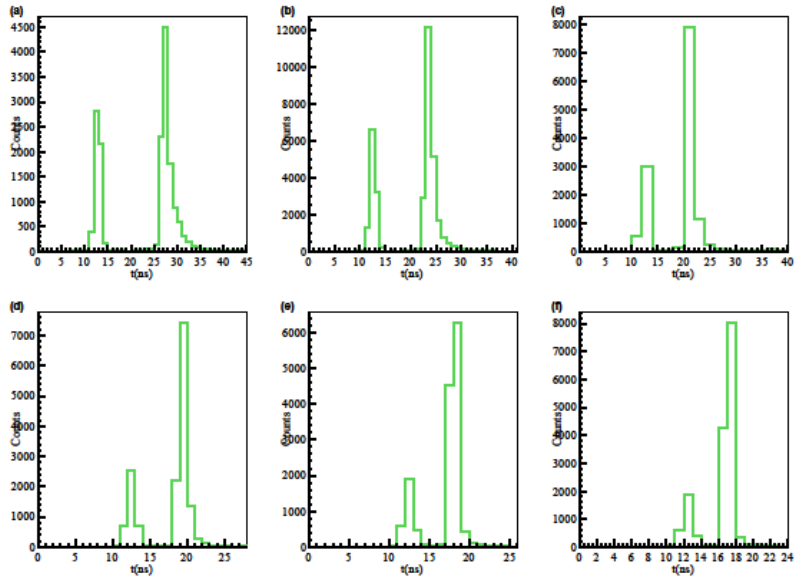

Fig.7 Flight times of $p/\pi^+$ at different momenta; the flight time of p is much more than that of $\pi^+$. (a) 500 MeV/c; (b) 600 MeV/c; (c) 700 MeV/c; (d) 800 MeV/c; (e) 900 MeV/c; (f) 1000 MeV/c.

Fig. 8 shows the amplitude spectra of p/$\pi^+$ at 500 MeV/c; the spectra can be fitted with a Landau function. The most possible value (MPV) of p shown in Fig. 8 is larger than that for $\pi+$. Fig. 9 shows the MPV of p/$\pi^+$ at the measured momenta; the amplitude of p decreases with increasing momentum, while the amplitude of $\pi^+$ is almost constant. The trend of energy deposition for p/$\pi^+$ at different momenta matches the Geant4 simulation results. After subtracting the baseline (180 mV), the amplitude of p is about 1-4 times higher than that of $\pi^+$.

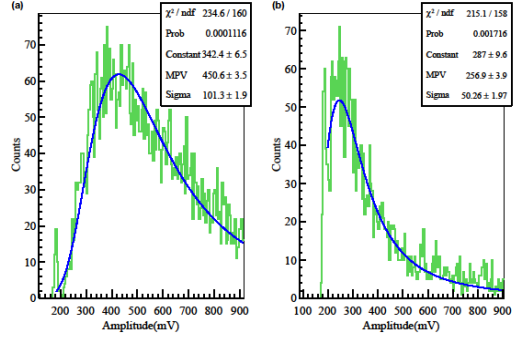

Fig.8 Amplitude spectra of p/$\pi^+$ at 500 MeV/c. (a) p; (b) $\pi^+$.

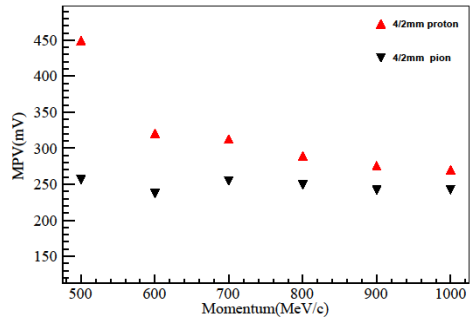

Fig.9. MPV of p/$\pi^+$ at different momenta

**3.3 Detection efficiency**

The detection efficiency of the THGEM detector is defined as the ratio of valid events to total events. On the basis of the threshold value (195 mV) set above, the detection efficiency of the THGEM detector for p and $\pi^+$ is calculated as shown in Fig. 10. The detection efficiency for p is slightly higher than that for $\pi^+$. Detection efficiency for p varies from 93% to 99%, and for $\pi^+$ from 82% to 88%.

According to Ref. [11] and the Geant4 simulation results, the deposition energy of p/$\pi^+$ at a given momentum should be different, as this is closely related to dE/dx. With increasing momentum, dE/dx for p decreases, while for $\pi^+$ there is only slight change. Considering fluctuation error and errors from experimental conditions, the trend of detection efficiencies for p/$\pi^+$ shown in Fig. 10 is reliable.

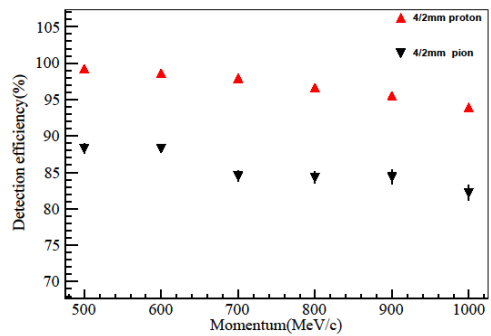

Fig.10. Detection efficiency for p/$\pi^+$ at different momenta

## 4 Conclusion and prospects

In future high energy experiments, the DHCAL will be an important component, especially for hadron jets etc. High spatial resolution, high counting rate and high detection efficiency will improve HCAL energy resolution. A HCAL based on THGEM detector should be competitive due to its above mentioned properties.

In this paper, we have made a preliminary study of the response of a THGEM detector to protons and $\pi^+$. A 4 mm drift gap and 2 mm induction gap THGEM detector was used, and worked rather stably in Ar+3% $iC_4H_{10}$ during the experiment, with gain of about 2000. The results show that detection efficiencies for p are slightly larger than for $\pi^+$, and the amplitude spectra of p/$\pi^+$ match the Geant4 simulation results. Based on the above results, the detection efficiency for $\pi^+$ at higher momentum should be better.

Certainly, it is important for a THGEM detector to improve efficiency for minimum ionizing particles (MIPs). In order to improve signal noise ratio, the working conditions of the THGEM detector should be optimized, including working gas and configuration. Meanwhile, for the possibility of use as a DHCAL sampling element, the THGEM detector should be as compact as possible. Further studies are ongoing.